\documentclass{mg11}

\newcommand{\be}{\begin{equation}}
\newcommand{\ee}{\end{equation}}
\newcommand{\bea}{\begin{eqnarray}}
\newcommand{\eea}{\end{eqnarray}}

\begin{document}



\title{PHANTOM DARK ENERGY AND ITS COSMOLOGICAL CONSEQUENCES}

\author{MARIUSZ P. D\c{A}BROWSKI}

\address{Institute of Physics, University of Szczecin,\\
Wielkopolska 15, 70-451 Szczecin, Poland\\
\email{mpdabfz@sus.univ.szczecin.pl}}


\begin{abstract}
I discuss the dark energy characterized by the violation of the null
energy condition ($\varrho + p \geq 0$), dubbed phantom. Amazingly, it is admitted by the
current astronomical data from supernovae. We discuss both
classical and quantum cosmological models with phantom as a source
of matter and present the phenomenon called phantom duality.
\end{abstract}

\bodymatter

\section{Introduction}\label{intro}

\noindent
Until a couple of years ago, the standard cosmological source of dark energy was
considered to be a slightly negative pressure matter ($- \varrho < p < 0$) with perhaps
time-evolving equation of state (quintessence), but not exceeding a ``mysterious'' barrier
$p = - \varrho = - \Lambda$, which corresponded to the cosmological
constant. Only the strong energy condition of Hawking and Penrose
($\varrho + 3p > 0$, $\varrho + p >0$) was presumably violated, and
the evolution of the universe in future contained two options: an
asymptotic emptiness or a Big-Crunch. However, a deeper analysis
of the data from supernovae, cosmic microwave background (WMAP) and
large-scale structure (SDSS) \cite{supernovaenew} shows that the dark
energy may also be the matter whose pressure is less than minus
the energy density and so violates the null energy condition, and
consequently, all the energy conditions. This matter is dubbed
phantom \cite{caldwell}, and it leads to qualitatively new types of
the evolution of the universe as a whole. I will
discuss these opportunities both in classical \cite{phantomcl} and quantum
\cite{phantomqu} cosmological context.

\section{Classical phantom cosmologies - Big-Rip and phantom duality}
\label{duality}

Phantom is dark energy of a strongly negative pressure which in
the easiest case may be simulated by a scalar field $\phi$ of negative
kinetic energy with the Lagrangian
\begin{equation}
L = \frac{l}{2} \partial_{\mu} \phi \partial^{\mu} \phi -
V(\phi)~,
\end{equation}
where $l=-1$ for phantom, $l=+1$ for standard scalar field, and $V(\phi)$
is the potential. In terms of the energy-momentum tensor for a
perfect fluid this gives ($\varrho$ - the energy density, $p$ - the
pressure): $\varrho = (l/2) \dot{\phi}^2 + V(\phi)$~,
$p = (l/2) \dot{\phi}^2 - V(\phi)$~,
and it surely violates the null energy condition $\varrho + p = l\dot{\phi}^2 >
0$, if $l=-1$. There are many other examples where phantom matter
appears naturally. For example, this is the case in Brans-Dicke
theory in the Einstein frame (provided the Brans-Dicke parameter $\omega < -
3/2$), in superstring cosmology, in brane cosmology, and in viscous
cosmology. Due to the energy conditions violation,
it makes a failure of the standard cosmic censorship conjectures,
black hole thermodynamics, positive mass theorems and other
renowned theorems of general relativity. It also leads to
classical and quantum instabilities \cite{instab}, which encourages some researchers
to disregard it as a serious candidate for the dark energy.
However, as it was already mentioned, its main
motivation does not come from the theoretical considerations.
On the contrary, it comes from the observational data. Of course its
stability is a problem, and there are suggestions how to
avoid that problem, too \cite{stabil}.


The most striking result which refers to phantom is that its
energy density $\varrho$ grows proportional to the scale factor
$a(t)$, i.e.,
\bea
\label{gammal0}
\varrho &\propto& a^{3\mid w+1 \mid} \hspace{0.5cm} {\rm for}
\hspace{0.5cm} w < -1 .
\eea
Then, unlike in a more intuitive standard matter case, where the growth of the
energy density corresponds to the decrease of the scale factor,
here, the growth of the energy density accompanies the expansion of the
Universe. This really gives a new scenario for the future evolution
of the universe, which has not been considered so far in
cosmology. There is a future singularity, which due to its peculiar
properties is called Big-Rip. On the approach to a
Big-Rip everything in the universe is pulled apart in a reverse
order - first clusters, then galaxies, solar systems, atoms,
nuclei etc. \cite{caldwell}.


One of the standard cosmological scenarios in cosmology is the
evolution from a Big-Bang to a Big-Crunch. In phantom cosmology
it is possible to start with a Big-Rip reach the minimum and
terminate at another Big-Rip. This is an example of the phantom
duality \cite{phantomcl} - a new symmetry of the field equations which allows to map
a large scale factor (cf. (\ref{gammal0})) onto a small one and vice versa due to a
change
\begin{equation}
a(t) \leftrightarrow 1/a(t) \hspace{0.5cm} {\rm or } \hspace{0.5cm} w+1 \leftrightarrow -(w+1)~.
\end{equation}
Similar symmetry was already discovered in the context of superstring
cosmology under the name of the scale factor duality
\cite{meissner} and further extended onto the brane cosmology
\cite{triality} (``phantom triality'').


Admission of phantom with $p < -\varrho$ enlarges possible set of
cosmological solutions. The most desirable are the solutions which
begin with a Big-Bang and terminate at a Big-Rip. This is because they
preserve all standard Hot-Big-Bang scenario results and agree with
the observational data, which suggests that there was a turning
point of the evolution of the universe (the standard matter
stopped dominating against phantom) just at the redshift $z=0.46$
\cite{supernovaenew}. However, as shown explicitly
\cite{phantomcl}, other interesting phantom cosmologies, due to its
strongly repulsive contribution to the dynamics appear. Among them
there are: Einstein Static Universe with two monotonic solutions -
one of them is monotonic towards a Big-Bang and another is monotonic
towards a Big-Rip; a monotonic solution which transits between the two Einstein
Static Universes (both ways); non-singular oscillating solutions etc.

\section{Quantum phantom cosmologies and the large-scale quantum effects}
\noindent
Due to the new types of classical trajectories it is also
interesting to ask about possible quantum cosmological
implications of phantom dark energy. Despite the large size of the universe
it is advisable to think of quantum effects in the region of Big-Rip
singularity since according to (\ref{gammal0}) the energy density
may reach the Planck scale of $10^{19}$ GeV there. In fact, it was
already shown \cite{kiefer} that there is a dispersion of the wave packets in
configuration space (with classical time coordinate eliminated)
at Big-Bang, Big-Crunch and at the maximum of expansion point in
recollapsing models. The question was studied whether such quantum
effects may also appear at a Big-Rip and at the minimum point of
expansion \cite{phantomqu}. It emerged that the answer is positive
for Big-Rip, but negative so far for the minimum point of
expansion. The latter might be due to the simple form of the
phantom potentials taken into account, and the question of such
quantum effects is still open. Besides, it seems reasonable to ask
if spreading of the wave packets in configuration space signaling
large-scale quantum effects may appear at the turning point of the
expansion - the one for $z=0.46$ at which standard matter loses
domination against the phantom. This is the matter for future
work.

\section{Discussion}\label{dicu}
\noindent
Apart from quintessence ($-\varrho < p <0$), global acceleration of
the universe gives strong observational motivation for a
non-standard phantom ($p<-\varrho$) type of matter as a candidate for dark
energy. Phantom dark energy may dominate the evolution of the
universe and additionally lead it to a Big-Rip singularity - a
state in which the whole matter pulled apart, though very dense. A Big-Rip
singularity may be dual to a Big-Bang/Big-Crunch singularity.
Finally, quantum effects may smear out Big-Rip singularity
due to large-scale quantum effects.

\section*{Acknowledgments}

This research has partially been supported by the
Polish Ministry of Science and Education grant No 1P03B 043 29 (years
2005-07).




\vfill

\end{document}